# An Agent-based Model of the Cognitive Mechanisms Underlying the Origins of Creative Cultural Evolution

**Liane Gabora**
Department of Psychology
University of British Columbia
Okanagan campus, 3333 University Way
Kelowna BC, V1V 1V7, CANADA
250-807-9849
liane.gabora@ubc.ca

**Maryam Saberi**
School of Interactive Art and Technology
Simon Fraser University
Surrey BC, V3T 0A3, CANADA
778-868-5159
msaberi@sfu.ca

## ABSTRACT

Human culture is uniquely cumulative and open-ended. Using a computational model of cultural evolution in which neural network based agents evolve ideas for actions through invention and imitation, we tested the hypothesis that this is due to the capacity for recursive recall. We compared runs in which agents were limited to single-step actions to runs in which they used recursive recall to chain simple actions into complex ones. Chaining resulted in higher cultural diversity, open-ended generation of novelty, and no ceiling on the mean fitness of actions. Both chaining and no-chaining runs exhibited convergence on optimal actions, but without chaining this set was static while with chaining it was ever-changing. Chaining increased the ability to capitalize on the capacity for learning. These findings show that the recursive recall hypothesis provides a computationally plausible explanation of why humans alone have evolved the cultural means to transform this planet.

### General Terms

Algorithms, Experimentation, Theory, Verification

### Author Keywords

EVOC; Homo erectus; action; agent-based model; creativity; cultural diversity; cultural evolution; gesture; innovation; invention; mimetic; origin of culture; recursive recall, self-triggered recall and rehearsal loop

## INTRODUCTION

How did humans come to be so creative? No other species has transformed this planet to anything near the degree that our species has. We now have the technology at our disposal to design computer models that enable us to investigate how our remarkable creativity came about. Humans are not only creative; we put our own spin on the inventions of others, such that new inventions build cumulatively on previous ones. This cumulative cultural change is referred to as the *ratchet effect* (Tomasello, Kruger, & Ratner, 1993), and it has been suggested that it is uniquely human (Donald, 1998).

This paper uses an agent-based simulation to address the question of what enabled humans to be creative enough that cumulative, open-ended cultural evolution could take hold and flourish. Our approach is motivated by the view that there is a distinct difference between (1) culture as a system involving transmission of chance innovations, which is observed in several species, and (2) culture as a process involving the accumulation and differentiation of strategic innovations, which is unique to our species. Only (2) constitutes a genuine evolutionary process, for to achieve descent with adaptive modification it is necessary that change be cumulative.

## A POSSIBLE MECHANISM FOR THE ORIGIN OF CULTURAL EVOLUTION

The minds of our earliest ancestors, *Homo habilis,* have been referred to as *episodic* because there is no evidence that their experience deviated from the present moment of concrete sensory perceptions (Donald, 1991). They were able to encode perceptions of events in memory, and recall them in the presence of a reminder or cue, but had little voluntary access to memories without environmental cues. They were therefore unable to voluntarily shape, modify, or practice skills and actions, and unable to invent or refine complex gestures or means of communicating.

*Homo erectus* lived between approximately 1.8 and 0.3 million years ago. The size of the *Homo erectus* brain was approximately 1,000 cc, about 25% larger than that of *Homo habilis*, at least twice as large as the brains of living great apes, and 75% the cranial capacity of modern humans (Aiello, 1996; Ruff et al., 1997). This period is widely referred to as the beginnings of cumulative culture. *Homo erectus* exhibited many indications of enhanced intelligence, creativity, and ability to adapt to their environment, including sophisticated, task-specific stone handaxes, complex stable seasonal home bases, and long-distance hunting strategies involving large game, and migration out of Africa.

This period marks the onset of the archaeological record and it is thought to be the beginnings of human culture. It is widely believed that this cultural transition reflects an underlying transition in cognitive or social abilities. Some have suggested that they owe their achievements to onset of *theory of mind* (Mithen, 1998) or the capacity to imitate (Dugatkin, 2001). However, there is evidence that other species possess theory of mind and the capacity to imitate (Heyes, 1998), yet do not compare to modern humans in intelligence and cultural complexity.

Evolutionary psychologists have suggested that the intelligence and cultural complexity of the *Homo* line can be attributed to the onset of *massive modularity* (Buss, 1999, 2004; Barkow, Cosmides, &Tooby, 1992). However, although the mind exhibits an intermediate degree of functional and anatomical modularity, neuroscience has not revealed vast numbers of hardwired, encapsulated, task-specific modules; indeed, the brain has been shown to be more highly subject to environmental influence than was previously believed (Buller, 2005; Byrne, 2000; Wexler, 2006).

Donald (1991) proposed that with the enlarged cranial capacity of *Homo erectus*, the human mind underwent the first of three transitions by which it—along with the cultural matrix in which it is embedded—evolved from the ancestral, pre-human condition. This transition is characterized by a shift from an *episodic* to a *mimetic mode* of cognitive functioning, made possible by onset of the capacity for voluntary retrieval of stored memories, independent of environmental cues. Donald refers to this as a *self-triggered recall and rehearsal loop*. Self-triggered recall enabled information to be processed recursively, and reprocessed with respect to different contexts or perspectives. Self-triggered recall allowed our ancestor to access memories voluntarily and thereby act out[1] events that occurred in the past or that might occur in the future. Thus not only could the mimetic mind temporarily escape the here and now, but by miming or gesture, it could communicate similar escapes in other minds. The capacity to mime thus ushered forth what is referred to as a *mimetic* form of cognition and brought about a transition to the mimetic stage of human culture. The self-triggered recall and rehearsal loop also enabled our ancestors to engage in a stream of thought. One thought or idea evokes another, revised version of it, which evokes yet another, and so forth recursively. In this way, attention is directed away from the external world toward one's internal model of it. Finally, self-triggered recall allowed for voluntary rehearsal and refinement of actions, enabling systematic evaluation and improvement of skills and motor acts.

[1] The term *mimetic* is derived from "mime," which means "to act out."

The recursive recall hypothesis is difficult to test directly, for if correct it would leave no detectable trace. It is, however, possible to computationally model how the onset of the capacity for recursive recall would affect the effectiveness, diversity, and open-endedness of ideas generated in an artificial society.

## THE MODEL

We tested Donald's hypothesis using an agent-based computational model of culture referred to as 'EVOlution of Culture', abbreviated EVOC (Gabora, 2008b, 2008c; Gabora & Leijnen, 2009; Leijnen & Gabora, 2009). EVOC has been used to address such questions as (1) how does the presence of leaders or barriers to the diffusion of ideas affect cultural evolution dynamics, (2) what is the effect of the ratio of creators to imitators in a society, and (3) what is the impact of locally clustering the creators. It is an elaboration of Meme and Variations, or MAV (Gabora, 1994, 1995), the earliest computer program to model culture as an evolutionary process in its own right (as opposed to modeling the interplay of cultural and biological evolution).The approach was inspired by Holland's (1975) genetic algorithm, or GA. The GA is a search technique that finds solutions to complex problems by generating a 'population' of candidate solutions through processes akin to mutation and recombination, selecting the best, and repeating until a satisfactory solution is found. The goal here was to distil the underlying logic of not biological evolution but cultural evolution, i.e. the process by which ideas adapt and build on one another in the minds of interacting individuals. In cultural evolution, the generation of novelty takes place through invention instead of through mutation and recombination as in biological evolution, and the differential replication of novelty takes place through imitation, instead of through reproduction with inheritance as in biological evolution (Gabora, 1996). EVOC (as did MAV) uses neural network based agents that could (1) invent new ideas by modifying previously learned ones, (2) evaluate ideas, (3) implement ideas as actions, and (4) imitate ideas implemented by neighbors. Agents do not evolve in a biological sense—they neither die nor have offspring—but do in a cultural sense, by generating and sharing ideas for actions. EVOC (like MAV) successfully models how 'descent with modification' occurs in a cultural context. The approach can thus be contrasted with computer models of how individual learning affects biological evolution (Best, 1999, 2006; Higgs, 2000; Hinton & Nowlan, 1987; Hutchins & Hazelhurst, 1991).

EVOC consists of an artificial society of neural network based agents in a two-dimensional grid-cell world. It is written in Joone, an object oriented programming environment, using an open source neural network library written in Java. We now summarize the key components of the agents and the world they inhabit.

### THE AGENTS

Agents consist of (1) a neural network, which encodes ideas for actions and detects trends in what constitutes a fit action, and (2) a body, which implements actions. The core of an agent is an auto associative neural network, as shown in Figure 1. The network learns ideas for actions. An idea is a pattern consisting of six elements that dictate the placement of the six body parts. Learning and training of the neural network is as per Gabora (1995).

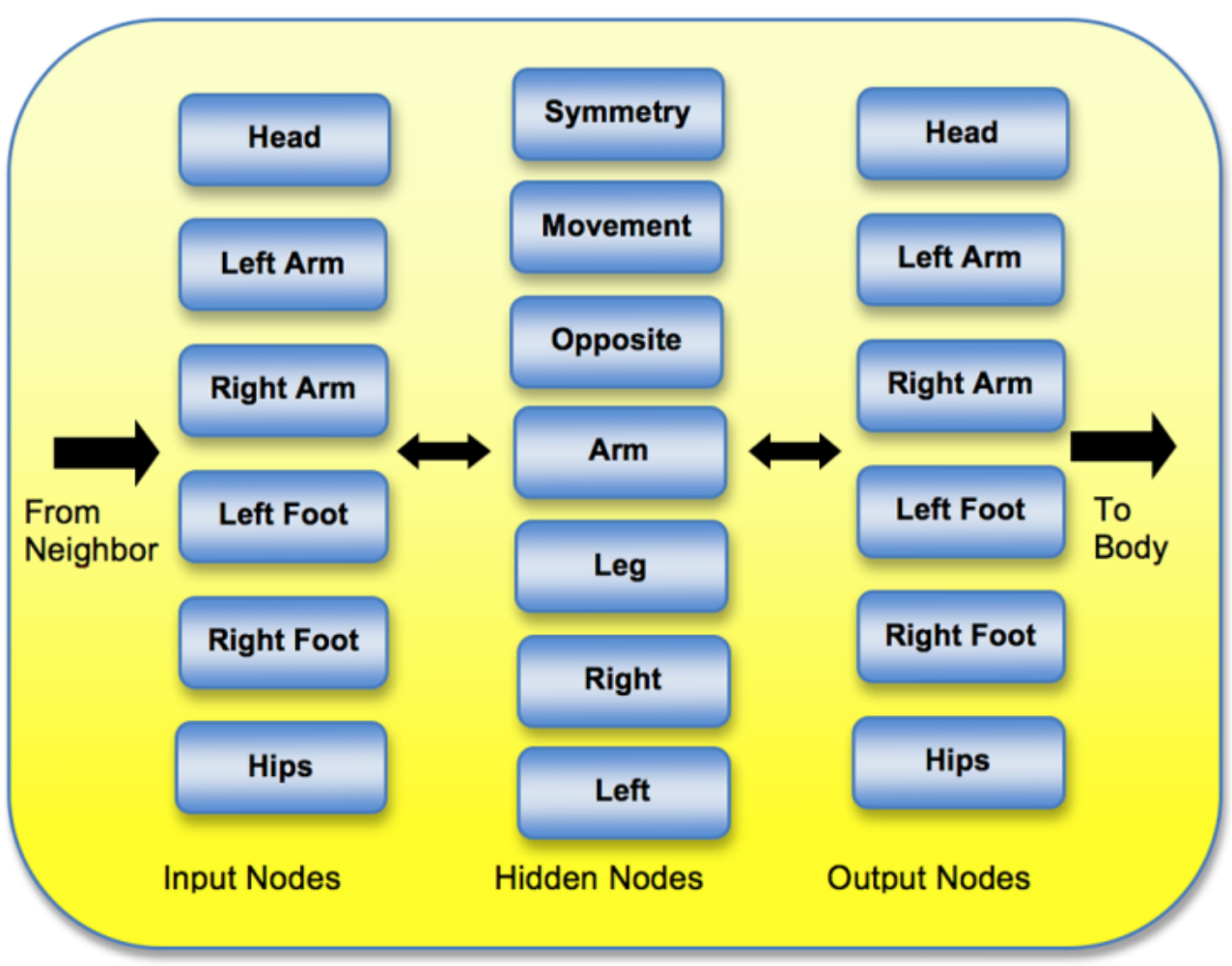

**Figure 1. The core of an agent is an autoassociative neural network composed of six input nodes and six corresponding output nodes that represent concepts of body parts (LEFT ARM, RIGHT ARM, LEFT LEG, RIGHT LEG, HEAD, and HIPS), and seven hidden nodes that represent more abstract concepts (LEFT, RIGHT, ARM, LEG, SYMMETRY, MOVEMENT and OPPOSITE). Input nodes and output nodes are connected to hidden nodes of which they are instances (e.g. RIGHT ARM is connected to RIGHT.) Activation of any input node activates the MOVEMENT hidden node. Same-direction activation of symmetrical input nodes (e.g. positive activation—which represents upward motion—of both arms) activates the SYMMETRY node. Sequential opposite movement of an ARM activates the OPPOSITE node. This hidden node was newly added to this version of the program to implement the chaining of actions to form multi-step actions. The hidden nodes are used to bias invention using learned trends about what constitutes a fit action.**

In EVOC, the neural network can also be turned off to compare results with a data structure that cannot detect trends, and thus invents ideas merely at random. If the fitness of an action is evaluated to be higher than that of any action learned thus far, it is copied from the input/output nodes of the neural network that represent concepts of body parts to a six digit array that contains representations of actual body parts, referred to as the body. Since it is useful to know how many agents are doing essentially the same thing, when node activations are translated into limb movement they are threshold such that there are only three possibilities for each limb: stationary, up, or down. Six limbs with three possible positions each gives a total of 729 possible actions. Only the action that is currently implemented by an agent's body can be observed and imitated by other agents.

### THE ARTIFICIAL WORLD

MAV allowed only worlds that were square and toroidal, or 'wrap-around' (such that agents at the left border that attempt to move further left appear on the right border). In EVOC the world can assume any shape, and be as sparsely or densely populated as required, with agents placed in any configuration. EVOC also allows for the creation of complete or semi-permeable permanent or eroding borders that decrease the probability of imitation along a frontier (although this was not used in the experiments reported here).

### INCORPORATION OF COGNITIVEPHENOMENA

The neural network can be turned off to compare neural network results with those obtained using a database that cannot learn trends, and thus invents ideas at random. The following cultural evolution parameters can also be turned off or on (in some cases to varying degrees).

#### Imitation

Fit actions diffuse through the society when agents copy neighbors' actions. During imitation, the input is the action implemented by a neighbor. The process of finding a neighbor to imitate works through a form of lazy (non-greedy) search. An imitating agent randomly scans its neighbors, and adopts the first action that is fitter than the action it is currently implementing. If it does not find a neighbor that is executing a fitter action than its own current action, it continues to execute the current action.

#### Invention

Agents generate new actions by modifying their initial action or an action that has been invented previously or acquired through imitation. During invention, the pattern of activation on the output nodes is fed back to the input nodes, and invention is biased according to the activations of the SYMMETRY, MOVEMENT, and OPPOSITE hidden nodes. (Were this not the case there would be no benefit to using a neural network.) The extent to which the generation of novelty is biased by past experience is referred to as the *learning rate*. To invent a new idea, for each node of the idea currently represented on the input layer of the neural network, the agent makes a probabilistic decision as to whether the position of that body part will change, and if it does, the direction of change is stochastically biased according to the learning rate. If the new idea has a higher fitness than the currently implemented idea, the agent learns and implements the action specified by that idea. (See Gabora, 1995 for further details.)

### Learning

Invention makes use of the ability to detect, learn, and respond adaptively to trends. Since a new action (or, in invention, a new idea for an action) is not learned unless it is fitter than the currently implemented action, new actions provide valuable information about what constitutes an effective idea. Thus knowledge acquired through the evaluation of previous actions is translated into educated guesses about what constitutes a successful action, and this is accomplished by updating the learning rate.

There are two learnable trends. The first concerns the overall level of activity involved in successful actions. Learning this trend can bias the generation of new ideas in favor of either more movement or less. Each body part starts out at a stationary rest position, and with an equal probability of changing to movement in one direction or the other. If the fitter action codes for more movement, increase the probability of movement of each body part. Do the opposite if the fitter action codes for less movement. If movement is generally beneficial, increase the probability that new actions involve movement of more body parts.

This trend is based on the assumption that movement in general (regardless of which particular body part is moving) can be beneficial or detrimental. This seems like a useful generalization since movement of any body part uses energy and increases the likelihood of being detected. It is implemented as follows:

$a_{m1}$ = movement node activation for current action

$a_{m2}$ = movement node activation for new action

$p(im)_I$ = probability of increased movement at body part $i$

$p(dm)_I$ = probability of decreased movement at body part $i$

IF ($a_{m2} > a_{m1}$)

THEN $p(im)_I = \mathrm{MAX}(1.0, p(im)_I + 0.1)$

ELSE IF ($a_{m2} < a_{m1}$)

THEN $p(im)_I = \mathrm{MIN}(0.0, p(im)_I - 0.1)$

$p(dm)_I = 1 - p(im)_i$

The second learnable trend is: if successful actions tend to be symmetrical (e.g. left arm moves to the right and right arm moves to the left), the probability increases that new actions are symmetrical. (See Gabora 1995 for details.) This generalization is biologically sensible, since many useful actions (e.g. walking) entail movement of limbs in opposite directions, while others (e.g. pushing) entail movement of limbs in the same direction. However, the reason for implementing a preference for symmetrical limb coordination is not to produce a biologically realistic model of motor control but to provide a systematic rationale for associating different actions with a graded distribution of fitness values.

The preference for symmetrical limb coordination is implemented in a manner analogous to that of the first rule. Of particular relevance to the current investigation, if multi-step actions involving one limb moving sequentially one direction and then the other are beneficial, the probability of inventing such actions increases.

In summary, each action is associated with a measure of its effectiveness, and generalizations about what seems to work and what does not are translated into guidelines for what action to take.

### Evaluation

Before committing to implementing an idea as an action, agents can assess how successful the action would be if it were implemented. They evaluate the effectiveness of their actions according to how well they satisfy needs using a pre-defined equation that rewards actions that make use of trends detected by the SYMMETRY, MOVEMENT, and OPPOSITE hidden nodes. A successful action is one in which all body parts except the head are moving, and limb movement is asymmetrical. (Thus if the left arm is moving up, the right arm is moving down, and vice versa.) For the experiments reported here with chaining turned off, where total body movement, $m$, is calculated by adding the number of active body parts, and $m_h$ is head movement, the fitness, $F_{nc}$, is calculated as follows:

$s_a$ = 1 if arms are moving symmetrically; 0 otherwise

$s_l$ = 1 if legs are moving symmetrically; 0 otherwise

$m_h$ = 1 if head is stationary; 0 otherwise

$F_{nc} = m + 1.5(s_a + s_l) + 2(1 - m_h)$

This gives a fitness function with a maximum value of ten, because while movement of the head contributes positively to the first term (which represents overall movement) it contributes negatively to the third term (which represents head stability). This corresponds to a relatively realistic action with respect to the need to attract the attention of, and maintain eye contact with others including a potential mate.

Since there are two optimal movements for the hips, one optimal movement for the head (or in this case, a lack of movement), two optimal arm combinations, and two optimal leg combinations, there are eight different optimal actions. In previous versions of the program that did not allow for chained actions, once an agent converged on one of the eight optimal actions (out of the possible 729 actions), it implemented that action for the rest of the run. Therefore, once all agents had found one or another of the optimal actions, the mean fitness of actions across the society plateaus, and the set of implemented actions became fixed.

### Chaining

Chaining, which is new to this version of the program, gives agents the opportunity to execute multi-step actions. For the experiments reported here with chaining turned on, if in the first step of an action an agent was moving at least one of its arms, it executes a second step, which again involves up to six body parts. If, in the first step, the agent moved one arm in one direction, and in the second step it moved the same arm in the opposite direction, it has the opportunity to execute a three-step action. And so on. The agent is allowed to execute an arbitrarily long action so long as it continues to move the same arm in the opposite direction to the direction it moved previously. Once it does not do so, the chained action comes to an end. The longer it moves, the higher the fitness of this multi-step chained action. Where $n$ is the number of chained actions, the fitness, $F_c$, is calculated as follows:

$$F_c = F_{nc} + (n - 1)$$

The fitness function with chaining provides a simple means of simulating the capacity for recursive recall.

We omit a detailed explanation of EVOC options that were not used in the experiments reported here, such as broadcasting, which allows the action of a leader to be visible to not just immediate neighbors, but all agents, thereby simulating the effects of media (Gabora, 2008a,b; Leijnen & Gabora, 2010).

## A TYPICAL RUN

Each iteration, every agent has the opportunity to (1) acquire an idea for a new action, either by imitation, copying a neighbor, or by invention, creating one anew, (2) update the knowledge-based operators, and (3) implement a new action. To invent a new idea, for each node of the idea currently represented on the input/output layer of the neural network, the agent makes a probabilistic decision as to whether change will take place, and if it does, the direction of change is stochastically biased by the knowledge-based operators. If the new idea has a higher fitness than the currently implemented idea, the agent learns and implements the action specified by that idea. To acquire an idea through imitation, an agent randomly chooses one of its neighbors, and evaluates the fitness of the action the neighbor is implementing. If its own action is fitter than that of the neighbor, it chooses another neighbor, until it has either observed all of its immediate neighbors, or found one with a fitter action. If no fitter action is found, the agent does nothing. Otherwise, the neighbor's action is copied to the input layer, learned, and implemented.

Fitness of actions starts out low because initially all agents are immobile. Soon some agent invents an action that has a higher fitness than doing nothing, and this action gets imitated, so fitness increases. Fitness increases further as other ideas get invented, assessed, implemented as actions, and spread through imitation. The diversity of actions initially increases due to the proliferation of new ideas, and then decreases as agents hone in on the fittest actions.

## THE GRAPHIC USER INTERFACE

The graphic user interface (GUI) makes use of the open source charting project, JFreeChart, enabling variables to be user defined at run time, and results to become visible as the program runs. The topmost output panel, shown in Figure 2, is provided to give the reader a sense of how the program works.

## EXPERIMENTS

In previous versions of EVOC, that did not allow chaining, there were only a fixed number of possible actions, and once all agents converged on optimal actions, cultural change ground to a halt. We hypothesized that if agents were able to chain actions together to form arbitrarily long actions, fitness would not stop increasing, and the agents would not converge on the same fixed set of actions.

For these experiments the artificial world was a toroidal lattice with 100 nodes, each occupied by a single, stationary agent. We used a von Neumann neighborhood structure, (agents only interacted with their four adjacent neighbors).

All graphs plot averages of ten runs, with an invention to imitation ratio of 1:1, and a rate of conceptual change of 1/6. (Since there are six body parts this means approximately one body part movement per step of an action). All agents were stationary at the start of a run.

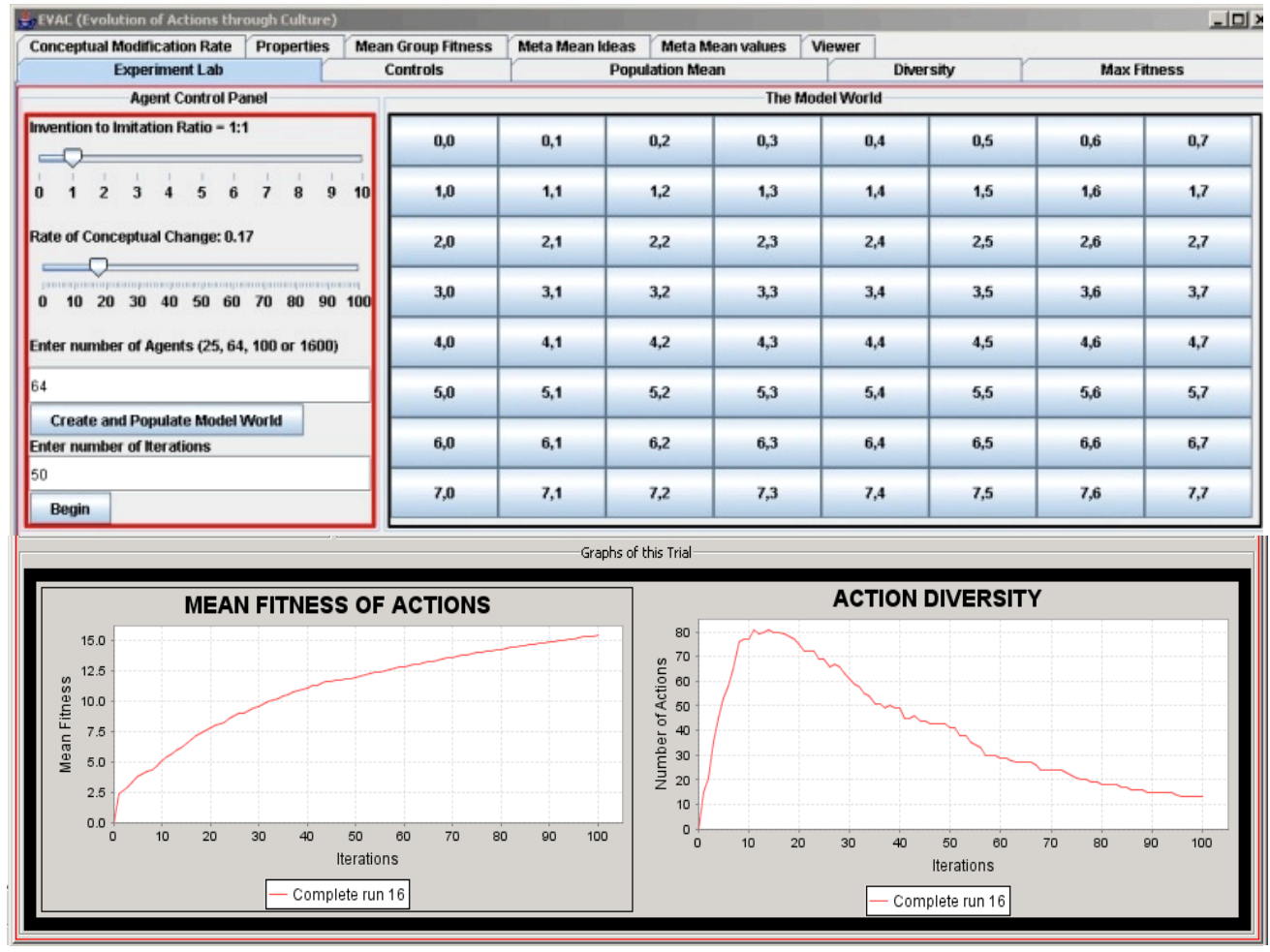

**Figure 2. Output panel of GUI. At the upper left one specifies the Invention to Imitation Ratio. This is the probability that a given agent, on a given iteration, invents a new idea for an action, versus the probability that it imitates a neighbor's action. Below it is Rate of Conceptual Change, where one specifies the degree to which a newly invented idea differs from the one it was based on. Below that is Number of Agents, which allows the user to specify the size of the artificial society. Below that is where one specifies Number of Iterations, i.e. the duration of a run. Agents can be accessed individually by clicking the appropriate cell in the grid on the upper right. This enables one to see such details as the action currently implemented by that agent, or the fitness of that action. The graphs at the bottom plot the mean idea fitness and diversity of ideas. Tabs shown at the top give access to other output panels.**

### Effect of Chaining on Fitness of Actions

Figure 3 shows the results of comparing how the capacity to recursively chain actions together affects the mean fitness of actions across the artificial society over the duration of a run.

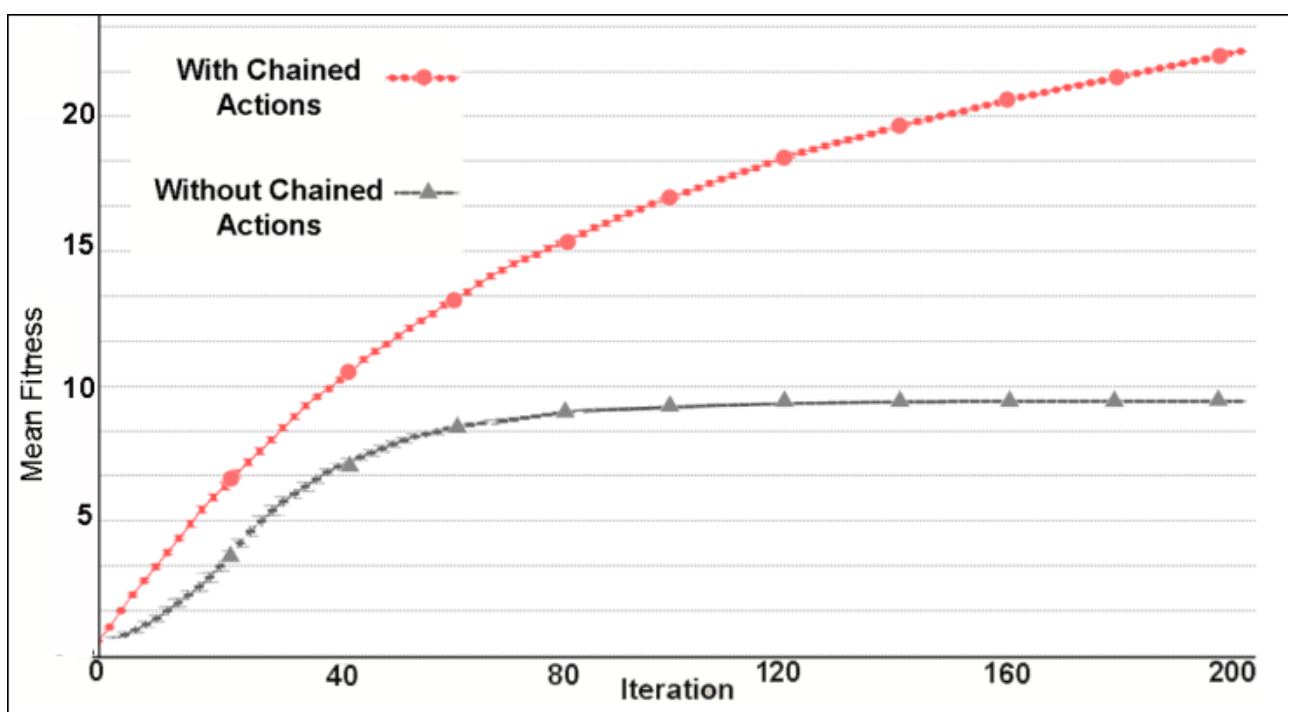

**Figure 3. Mean fitness of actions in the artificial society with chaining versus without chaining.**

The capacity to chain together simple actions to form more complex ones increases the mean fitness of actions across the artificial society. This is most evident in the later phase of a run. Without chaining, agents converge on optimal actions, and the mean fitness of action reaches a plateau. With chaining, however, there is no ceiling on the mean fitness of actions. By the 100th iteration it reached almost 15, indicating a high incidence of chaining.

### Effect of Chaining on Diversity of Actions

Figure 3 shows the results of comparing how the capacity to recursively chain actions together affects the mean fitness of actions across the artificial society over the duration of a run.

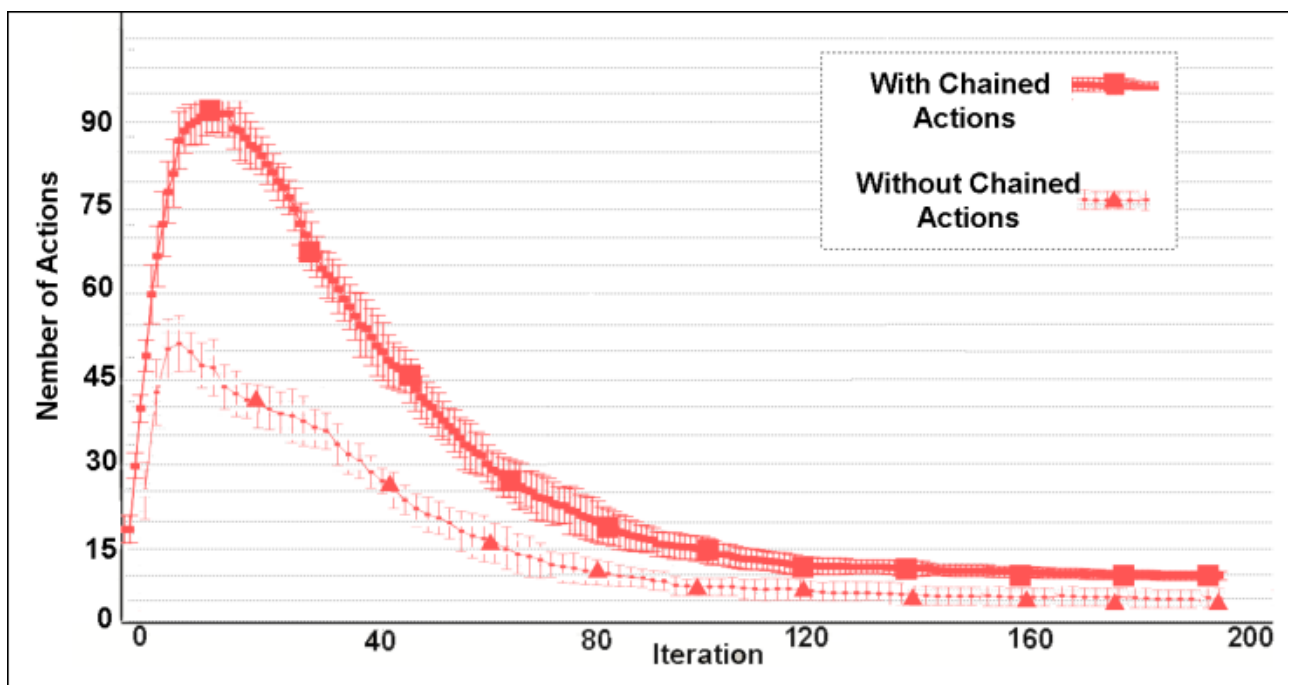

**Figure 4. Mean number of different actions in the artificial society with chaining (continuous line) versus without chaining (dashed line).**

Chaining also increases the diversity of actions. This is most evident in the early phase of a run before agents begin to converge on optimal actions, though it remains the case throughout the duration of a run. Interestingly, although in both cases there is convergence on optimal actions, without chained actions, this is a static set (thus mean fitness plateaus) whereas with chained actions the set of optimal actions is always changing, as increasingly fit actions are found (thus mean fitness keeps increasing).

### The Effect of Chaining on Learning

Recall that agents have the capacity to learn trends from past experiences (using the knowledge-based operators), and thereby bias the generation of novelty in directions that have a greater than chance probability of being fruitful. Results obtained previously with this kind of computer model of cultural evolution showed that learning increases the mean fitness of actions throughout the duration of a run (Gabora, 1995). Since chaining provides more opportunities to capitalize on the capacity to learn, we hypothesized that chaining would accentuate the impact of learning on the mean fitness of actions across the artificial society. Figure 5 compares the effect of chaining with knowledge-based operators turned on (agents are able to learn) versus when they are turned off (agents are unable to learn).

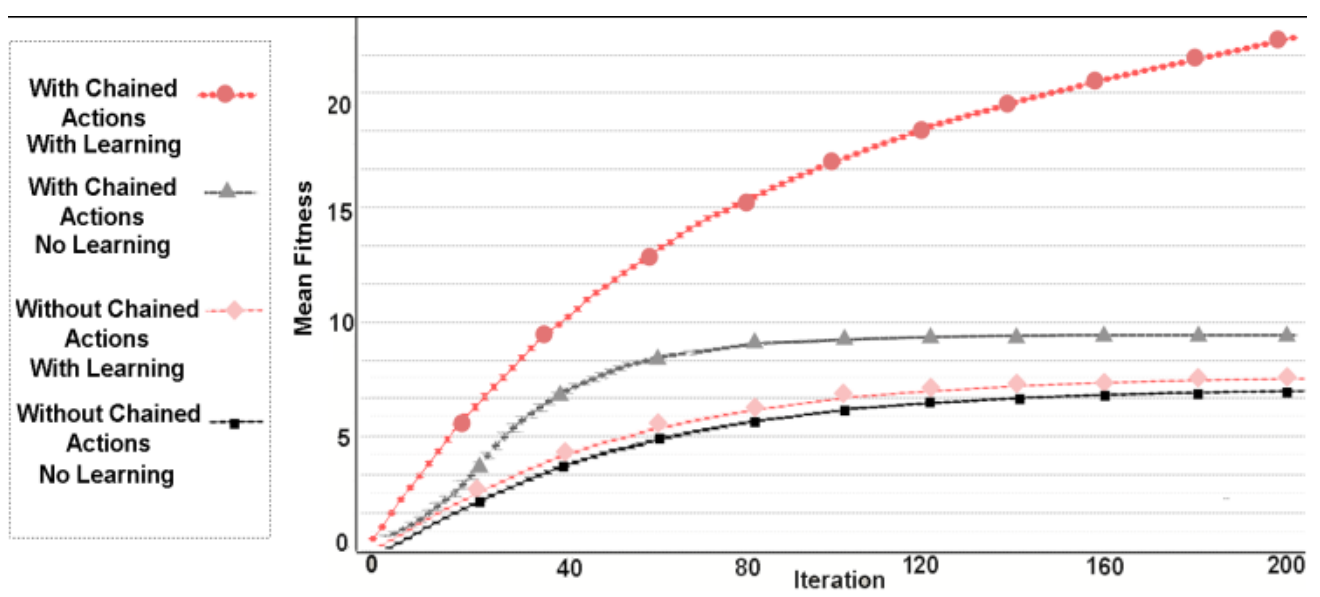

**Figure 5. Mean fitness of actions in the artificial society with both chained actions and learning (top), with just chained actions (second from the top) with just learning (second from the bottom), and with neither chained actions nor learning (bottom).**

This hypothesis was confirmed. Since waving an arm back and forth increased activation of the OPPOSITE hidden node, agents learned to generate increasingly longer moves throughout the duration of a run. When agents are given the capacity to chain together simple actions to form more complex ones, the increase in mean fitness of actions throughout the duration of a run is more pronounced, indicating that the society as a whole benefits much more from learning.

### Discussion

Many species possess what could be called culture, but human culture is unique in that ideas build on each other cumulatively; indeed culture is widely thought to be an evolutionary process (e.g. Bentley, Ormerod, & Batty, 2011; Cavalli-Sforza & Feldman, 1981; Gabora, 1996, 2008; Hartley, 2009;Mesoudi, Whiten & Laland, 2004, 2006; Whiten, Hinde, Laland, & Stringer, 2011). Using an agent based computer model of cultural evolution, we obtained support for the hypothesis that the onset of cumulative, open-ended cultural evolution can be attributed to the evolution of a self-triggered recall and rehearsal loop, enabling the recursive chaining of thoughts and actions.

Our results suggest that it is the capacity for recursive recall that makes possible the open-endedness of some computational models of language evolution (*e.g.* Kirby, 2001). Note that in the chaining versus no chaining conditions the size of the neural network is the same, but how it is *used* differs. This suggests that it was not larger brain size *per se* that initiated the onset of cumulative culture, but that larger brain size enabled episodes to be encoded in more detail, allowing more routes for reminding and recall, thereby facilitating the ability to recursively re-describe information stored in memory (Karmiloff-Smith, 1992), thereby tailor it to the situation at hand.

By demonstrating that onset of the capacity for recursive recall constitutes a viable explanation for how cultural change became cumulative and open-ended, we believe that this research is a pivotal step forward in the development of a scientific theory of cultural evolution. Nevertheless, the model is simplified, and caution must be taken applying such results to the real world. The agents can be said to have the goal of optimizing the fitness of their actions, but they do not have values, norms, or social bonds remotely like ours. The goal behind a computational model such as EVOC, however, is to extract the skeleton of a process as complex as cultural evolution and investigate questions that are difficult to investigate by other means. To our knowledge, this is the only experimental test of, and tentative support for, the hypothesis that the capacity for self-triggered recall of thoughts and recursive execution of actions played a key role in ratcheting the kind of cumulative, open-ended novelty and cultural complexity that surrounds us. Despite that evolved cultural complexity essentially defines our humanness, we know of no other computational modeling support for this or any other competing hypothesis concerning how the transition from episodic to mimetic cognition came about.

Future work will focus on simulating more complex forms of recursive recall that would require a larger neural network, and corroborating findings obtained with the model with archaeological and anthropological evidence. Another direction of future research involves using EVOC to investigate competing hypotheses concerning what cognitive mechanism could have given rise to the explosive cultural transition in the Middle-Upper Paleolithic (Klein, 1999; Leaky, 1984).

## ACKNOWLEDGMENTS

We would like to acknowledge grants to the first author from the *Social Sciences and Humanities Research Council of Canada* and the Concerted Research Program of the *Fund for Scientific Research, Belgium*.